\documentstyle[epsfig]{aipproc}
\begin{document}
\title{A Relativistic Quark Model for Baryons}

\author{Bernard Metsch}
\address{Institut f\"ur Theoretische Kernphysik\\
Universit\"at Bonn\\
Nu{\ss}allee 14-16, D53115 Bonn, Germany}

\maketitle

\begin{abstract}
  We present a relativistic quark model for baryons, based on the
  Bethe--Salpeter equation in instantaneous approximation. Confinement
  is implemented by an interaction kernel which essentially is a
  linearly rising potential with a spin-dependence chosen such as to
  minimize spin-orbit effects. The fine structure of the baryon
  spectrum follows from an effective quark-interaction based on
  instanton effects. Results for the spectra of all baryons build from
  $u,d,s$--quarks are presented. In particular it is found, that the
  present relativistic setup can account for the low position of
  Roper--like resonances in all sectors.
\end{abstract}

Despite many efforts solving the bound state problem of QCD a
consistent, quantitative description of mesons and baryons is still
one of the major challenges in elementary particle physics. If one
aims at a comprehensive description of hadronic excitations the most
successful approach is doubtless the constituent quark model, where
the gluonic degrees of freedom are eliminated in favor of constituent
quarks with effective masses and quark interaction potentials. In its
non-relativistic version this approach has been most extensively
studied by Isgur and Karl, see e.g.~\cite{Isg78}. Later on this has been
improved, by including relativistic corrections both to the kinetic
energy and to the quark potentials. In spite of the success in
describing properties of baryons this procedure seems questionable in
view of the high velocities of the light quarks and the large
''binding energies'' one finds especially for mesons like the pion.
Furthermore the lack of formal covariance impedes a consistent
calculation of electromagnetic properties at large momentum transfers,
such as electromagnetic form factors.

It is the aim of the present contribution to extend a covariant quark
model, which we developed for mesons on the basis of the
Bethe-Salpeter equation (see~\cite{Met97} and references therein) to
baryons. The Bethe--Salpeter amplitude for a three-fermion bound state
with 4-momentum $\bar{P}$ defined by $ \chi_{\bar P}(x_1,x_2,x_3)
\equiv \langle 0 \left | T \, \Psi^1(x_1) \Psi^2(x_2) \Psi^3(x_3)
\right| \bar{P}\rangle $ satisfies the homogeneous Bethe Salpeter
equation, which after Fourier transformation in momentum space and
introduction of the Jacobi-momenta $p_\xi,p_\eta$ reads:
\begin{eqnarray}
\label{eq:1}
\lefteqn{\!\!\chi_{\bar P}(p_\xi,p_\eta) =
-i S^F_{1}(p_1) S^F_{2}(p_2) S^F_{3}(p_3)
\int\!\!\frac{d^4\!p_\xi'}{(2\pi)^4}\;\frac{d^4\!p_\eta'}{(2\pi)^4}\;
K^{(3)}(\bar P,p_\xi,p_\eta,p_\xi',p_\eta')\;
\chi_{\bar P}(p_\xi',p_\eta')}
\nonumber \\
& & -i \Bigl[S^F_{1}(p_1)S^F_{2}(p_2)
\int\!\!\frac{d^4\!p'_{\xi }}{(2\pi)^4}\;
K^{(2)}((p_1+p_2,p_{\xi},p'_{\xi})
\chi_{\bar P}(p_\xi',p_\eta) + \mbox{ 2 cycl. perm.}\Bigr]\,,
\end{eqnarray}
where $S^F_i$ are the full single quark propagators,
$p_i(p_{\xi},p_{\eta})$ are single particle 4-momenta depending on the
Jacobi-momenta, and $K^{(2)}$ and $K^{(3)}$ are the two-particle and
three-particle irreducible interaction kernels, respectively.

In the spirit of the covariant meson quark model~\cite{Met97} we will
make the following simplifying assumptions, which nevertheless still
respect relativistic covariance: The instantaneous approximation can be
readily formulated if we keep only the first term on the right hand
side of eq. (\ref{eq:1}), assume that the full propagators are given
by their free form $S^F_i(p) = i/(p\!\!\!/ - m_i+i\varepsilon)$, with
an effective constituent quark mass $m_i$ and suppose that that the
three particle kernel $K^{(3)}$ is instantaneous in the rest frame of
the bound state: $ K^{(3)}(\bar P, p_\xi, p_\eta, p_\xi', p_\eta') =
V(\vec{p}_{\xi},\vec{p}_{\eta}, \vec{p}_{\xi}',\vec{p}_{\eta}'), $
independent of $p^0_\xi,p^0_\eta$. With these approximations it is
possible to perform the integrations over $p^0_\xi$ and $p^0_\eta$ in
eq.~\ref{eq:1} analytically by the residue theorem and reduce the
Bethe--Salpeter equation to the 3-fermion (full) Salpeter equation
\begin{eqnarray}\label{eq:2}
  \Phi(\vec{p}_\xi,\vec{p}_\eta)
  &=&\bigg[\frac{\Lambda_1^+(\vec{p}_1)
          \otimes\Lambda_2^+(\vec{p}_2)
          \otimes\Lambda_3^+(\vec{p}_3)}{
            M-\omega_1-\omega_2-\omega_3\;+\;i\epsilon}
          +\frac{\Lambda_1^-(\vec{p}_1)
          \otimes\Lambda_2^-(\vec{p}_2)
          \otimes\Lambda_3^-(\vec{p}_3)}{
            M+\omega_1+\omega_2+\omega_3\;-\;i\epsilon}
     \bigg]\gamma^0\otimes\gamma^0\otimes\gamma^0\nonumber \\
  & &\times\;
  \int\frac{d^3\!p_\xi'}{(2\pi)^3}\;\frac{d^3\!p_\eta'}{(2\pi)^3}\;
  V(\vec{p}_\xi,\vec{p}_\eta,\vec{p}_\xi',\vec{p}_\eta')
  \Phi(\vec{p}_\xi',\vec{p}_\eta')
\end{eqnarray}
for the Salpeter-Amplitude
$
  \Phi(\vec{p}_\xi,\vec{p}_\eta)
  \equiv
  \int \frac{dp_\xi^0}{2\pi}\;\frac{dp_\eta^0}{2\pi}\;\chi_{\bar P}
  \left((p_\xi^0,\vec{p}_\xi),(p_\eta^0,\vec{p}_\eta)\right)
  \mid_{\bar P=(M,0)}\;.
$
Here we introduced the abbreviations: $ \omega_i = \omega_i(\vec{p}_i)
= \sqrt{m_i^2+|\vec{p}_i|^2} $ with $
\Lambda_i^{\pm}=\Lambda_i^{\pm}(\vec{p}_i)\equiv{(\omega_i\pm
  H_i(\vec{p}_i))}/{2\omega_i}$, where $
H_i(\vec{p}_i)\equiv\gamma^0\;(\vec{\gamma}\!\cdot\!\vec
p_i+m_i)=\vec\alpha\!\cdot\!\vec{p}_i+\beta\; m_i$ is the single
particle Dirac Hamiltonian.

Using $H_i(\vec{p}_i)\;\Lambda_i^{\pm}(\vec{p}_i)=\pm \omega_i\;
\Lambda_i^{\pm}$ we can write the Salpeter equation in the concise
form
\begin{equation}\label{eq:4}
  ({\cal H}\Phi)(\vec{p}_\xi,\vec{p}_\eta) = 
  M \Phi(\vec{p}_\xi,\vec{p}_\eta),
\end{equation}
where we define the Salpeter Hamiltonian ${\cal H}$ through
\begin{eqnarray}\label{eq:5}
  ({\cal H}\Phi)(\vec{p}_\xi,\vec{p}_\eta)
  &=& \sum_{i=1}^3 H_i(\vec{p}_i)\Phi(\vec{p}_\xi,\vec{p}_\eta)
  +
  \left(
    \Lambda^{+}_1\otimes  \Lambda^{+}_2\otimes \Lambda^{+}_3
  + \Lambda^{-}_1\otimes  \Lambda^{-}_2\otimes \Lambda^{-}_3\right) 
  \nonumber \\
  & & \gamma^0\otimes \gamma^0\otimes \gamma^0 \;
  \int\frac{d^3\!p_\xi'}{(2\pi)^3}\;\frac{d^3\!p_\eta'}{(2\pi)^3}\;
  V(\vec{p}_\xi,\vec{p}_\eta,\vec{p}_\xi',\vec{p}_\eta')\;
  \Phi(\vec{p}_\xi',\vec{p}_\eta').
\end{eqnarray}
We require, that the interaction kernel respects
parity and time reversal invariance and that the Salpeter Hamiltonian
${\cal H}$ is selfadjoint with respect to a scalar product following
from the normalization condition on the amplitudes. However, ${\cal H}$
is not positive definite, and accordingly ${\cal H}$ possesses
positive and negative eigenvalues. From CPT-invariance it can be
shown, that if $\Phi$ is solution with eigenvalue $-M$, i.e.  $ ({\cal
  H}\Phi)(\vec{p}_\xi,\vec{p}_\eta) = -M \Phi
(\vec{p}_\xi,\vec{p}_\eta) $ then $\tilde \Phi \equiv \otimes_{i=1}^3
\gamma^0_i\gamma^5_i\Phi$ is a solution of the Salpeter equation with
eigenvalue $+M$: $ ({\cal H}\tilde\Phi)(\vec{p}_\xi,\vec{p}_\eta) = M
\tilde\Phi (\vec{p}_\xi,\vec{p}_\eta) $ and the solution $\tilde\Phi$
has a parity opposite to $\Phi$. This allows the interpretation of the
negative energy solutions for a given set of quantum numbers as
antibaryon states: After the transformation $\Phi \to \tilde \Phi$
these yield solutions with the same quantum numbers, but of opposite
parity. In this manner we thus find in the present approach the {\em same}
number of states as in the nonrelativistic quark model. In fact this
is an {\em a posteriori} justification for ignoring the second term on
the right hand side of eq. (\ref{eq:1}): This term would, even in
instantaneous approximation mix  the amplitudes $\Phi^{+++}$ and
$\Phi^{---}$ considered here with amplitudes such as $\Phi^{++-}$,
which contain both positive- and negative energy spinors. There is,
however, no empirical evidence, that such amplitudes are relevant for the
spectrum of baryon states, which is explained excellently by the naive
nonrelativistic model.

The Salpeter equation (\ref{eq:4}) is solved numerically by
diagonalization of ${\cal H}$ in (\ref{eq:5}) in an appropriate finite
basis of amplitudes with conserved quantum numbers and definite
permutational symmetry, which, due to the special structure of the
projectors in (\ref{eq:5}) can be constructed as in the
nonrelativistic model. Analogous to the meson model~\cite{Met97} we
parameterized V in~\ref{eq:2} as a sum of a confinement interaction
\begin{equation}
  \label{eq:6}
  {\cal V}(\vec x_1,\vec x_2,\vec x_3) = 
  \frac{3}{4} a 
  \left[1+ \sum_{i<j=1}^3 \gamma^0_i \gamma^0_j
  \right] \nonumber \\
+ \frac{1}{2} b \sum_{i<j=1}^3 |\vec x_i - \vec x_j| \left[ 1 + 
  \gamma^0_i \gamma^0_j \right]
\end{equation}
with offset $a$ and string tension $b$ and an instanton induced interaction
\begin{equation}
  \label{eq:7}
  {\cal W}(\vec x_1,\vec x_2,\vec x_3) = \sum_{i<j=1}^3
  -4 \left[ g {\cal P}^f_{{\cal A}_{ij}(nn)} +
 g' {\cal P}^f_{{\cal A}_{ij}(ns)} \right] \left[ 1 +
  \gamma^5_i\gamma^5_j\right] 
  N_\lambda e^{-(\vec x_i-\vec x_j)^2/\lambda^2}\,,
\end{equation}
where ${\cal P}_{\cal A}^f$ projects on flavour antisymmetric pairs
with strength $g$ and $g'$, and the original contact interaction was
regularized by a Gaussian with range $\lambda$.

\begin{table}[!ht]
        \caption{Comparison of experimental [3] and calculated masses (in
          MeV) of selected $\Delta$ resonances.}
        \label{tab:1}
        \begin{tabular}{rclrclrclrcl}
        $J^{\pi}$&Exp.&Cal.&
        $J^{\pi}$&Exp.&Cal.&
        $J^{\pi}$&Exp.&Cal.\\
        \tableline
        $\frac{1}{2}^+$&1700-1800&1760& 
        $\frac{7}{2}^+$&1940 1960&1915& 
        $\frac{13}{2}^+$&        &2665\\
        $\frac{1}{2}^-$&1615-1675&1600&
        $\frac{7}{2}^-$&2155-2360&2080&
        $\frac{13}{2}^-$&2550-2870&2580\\
        $\frac{3}{2}^+$&1230-1240&1260& 
        $\frac{9}{2}^+$&2140-2550&2290&
        $\frac{15}{2}^+$&2750-3190&2755\\
        $\frac{3}{2}^-$&1670-1770&1570&
        $\frac{9}{2}^-$&2100-2500&2180&
        $\frac{15}{2}^-$&        &2840\\
        $\frac{5}{2}^+$&1870-1920&1840& 
        $\frac{11}{2}^+$&2300-2500&2375&
        & & \\
        $\frac{5}{2}^-$&1920-1970&2015&
        $\frac{11}{2}^-$&        &2490&
        & & \\

 \end{tabular}
\end{table}

\begin{table}[!ht]
        \caption{Comparison of experimental [3] and calculated masses
        (in MeV) of  $N, \Lambda, \Sigma $ and $\Xi$ resonances.}
        \label{tab:2}
        \begin{tabular}{rclclclclcl}
          \multicolumn{1}{c}{} &
          \multicolumn{2}{c}{$N$} &
          \multicolumn{2}{c}{$\Lambda$} & 
          \multicolumn{2}{c}{$\Sigma$} & 
          \multicolumn{2}{c}{$\Xi$} \\
        \tableline
        $J^{\pi}$ & Exp.& Cal.& Exp.& Cal..& Exp.& Cal.& Exp.& Cal.\\
        \tableline
        $\frac{1}{2}^+$
&939       &  935 & 1116      & 1110 & 1189-1198 & 1210 & 1313-1315 & 1270 \\
        $\frac{1}{2}^+$
&1430-1470 & 1470 & 1560-1700 & 1585 & 1630-1690 & 1760 & 1950? & 1965\\
        $\frac{1}{2}^+$
&1680-1740 & 1750 & 1750-1850 & 1775 & 1738-1772 & 1898 &      & 2095\\
        $\frac{1}{2}^-$
&1520-1555 & 1470 & 1402-1410 & 1450 & 1600-1633 & 1610 & 1690?& 1710 \\
        $\frac{1}{2}^-$
&1640-1680 & 1625 & 1660-1680 & 1650 & 1730-1800 & 1735 &      & 1910 \\
        $\frac{3}{2}^+$
&1650-1750 & 1735 & 1850-1910 & 1835 & 1375-1400 & 1415 & 1529-1533 &1570 \\
        $\frac{3}{2}^-$
&1515-1530 & 1420 & 1518-1522 & 1515 & 1580-1583 & 1675 & 1819-1830 & 1795\\

 \end{tabular}
\end{table}

Since the instanton induced interaction acts only on states, which
contain a flavour antisymmetric pair, the flavour decuplet states are
determined by the confinement potential alone: As can be seen from
table~\ref{tab:1} our present {\em Ansatz}, which minimizes spin-orbit
effects can indeed account reasonably well for the $\Delta$-resonances
up to spin $\frac{15}{2}^+$. The instanton force does determine the
spectra of other sectors and the mass spectrum resulting from our
model for lower lying states is quite remarkable: Usually, in
non-relativistic quark models (both with and without relativistic
corrections) the Roper resonance and its strange partners appear much
too high in comparison with the experimentally determined resonance
positions and are lowered only when additional interactions (commonly
meson dressing effects) are taken into account.  In our covariant
model this is no longer the case; once the model parameters are fixed
to reproduce the correct ground state baryon masses, the spectrum of
the lowest excitations of positive parity is already well described,
see table~\ref{tab:2}.  Also the conspicuous low position of the
$\Lambda(1405;\frac{1}{2}^-)$-resonance is explained through a
combination of relativistic and instanton effects. We consider these
results to be a promising starting point for further calculations
which will also involve the computation of electro-weak observables.

The contributions from Klaus Kretzschmar, Ulrich L\"oring, Ralf Ricken
and Herbert Petry are gratefully acknowledged.

\end{document}